\documentclass{kluwer_mod}    

\usepackage[dvips]{graphicx}

\newdisplay{guess}{Conjecture}

\begin{document}                                                          

\begin{article}
\begin{opening}         
\title{Galactic Cannibalism: \\
  The Origin of the Magellanic Stream
  \thanks{This work was partly funded by the Victorian Partnership
for Advanced Computing, through their Expertise Grant.}}
\author{S.~T.\surname{ Maddison}}
\author{D.\surname{ Kawata}}
\author{B.~K.\surname{ Gibson}}
\runningauthor{Maddison, Kawata \& Gibson}
\runningtitle{Galactic Cannibalism}
\institute{Centre for Astrophysics and Supercomputing, 
           Swinburne University, Australia}


\end{opening}  

\section{Background}  

\vspace{-4.0mm}
The aim of this work is to model the formation of the Magellanic Stream via 
the resulting tidal gravitational field from the merger of the Milky Way 
with the Large and Small Magellanic Clouds (LMC and SMC, respectively).
Two popular, yet competing, scenarios for the Stream's formation are based
upon tidal disruption (Lin \& Lynden-Bell 1977; Gardiner \& Noguchi, 1996)
or ram-pressure stripping (Moore \& Davis 1994).  
The recent discovery of the Leading Arm Feature (Putman et~al. 1998) has
strengthened the case for tidal disruption, in which both trailing and
leading gas streams are a natural outcome.

\vspace{-5.0mm}
\section{Preliminary Simulations}

\vspace{-4.0mm}
To simulate the merger of the Milky Way--LMC--SMC system, we used the 
\texttt{TreeSPH} 
code of Kawata (2001) which includes a self-consistent treatment of
self-gravity, gas dynamics, radiative cooling, 
star formation, supernova feedback, and metal enrichment.  
The initial conditions for each galaxy was constructed using 
\texttt{GalactICs} (Kuijken \& Dubinski 1995).
Starting with the current positions of the Clouds, orbits were
integrated backwards in time (similar to Murai \& Fujimoto 1980),
resulting in appropriate initial conditions.  We then traced the system's
evolution from time T=$-$2~Gyr to the present.

\begin{center}
\begin{figure}
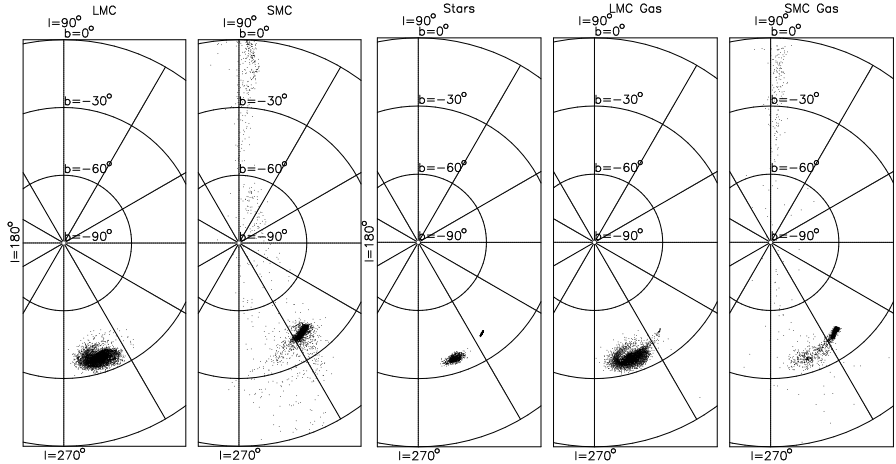
 
 \centering
 \leavevmode
 \includegraphics[width={.4\columnwidth}]{nbody.epsi}%
 \includegraphics[angle=-90,origin=br,width={.6\columnwidth}]{sc01.epsi}%
\caption[]{The left two panels show the present-day N-body only results for our
3-galaxy merger; the LMC and SMC particles are shown.
The three right panels show the results of 
the full hydrodynamics 3-galaxy merger model with (from left to right) the
present-day positions of the
stars, the LMC gas, and the SMC gas.}
\label{fig-MS}
\end{figure} 
\end{center}

We next compared the results of our pure N-body mergers 
with those of our full hydrodynamics simulations (including
star formation, cooling and supernova feedback).  In the N-body only 
case, we found that material from the SMC was tidally stripped, resulting
in the formation of the Magellanic Stream and an associated 
Leading Arm (see left two panels of Figure~\ref{fig-MS}).  

However, observations show that the Stream is apparently devoid of stars,
comprised primarily of gas (e.g. Br\"uck \& Hawkins 1983; Ostheimer et~al.
1997).  In our full hydrodynamics simulations (with star 
formation), the SMC was again severely disrupted.  In this case though
we found that the Stream contained only gas stripped from the SMC, with no 
accompanying stars (see right three panels of Figure~\ref{fig-MS}).
The outer tenuous material of the SMC -- gas -- was stripped to produce the 
Stream in both cases.  

\vspace{-17.0mm}
\section{Discussion}

\vspace{-4.0mm}
Preliminary hydrodynamical and N-body simulations were undertaken 
with self-consistent star formation and gas heating/cooling.  
Our models successfully 
recover a pure gas Magellanic Stream, similar to that observed.
Self-consistent treatments of star formation histories of the LMC and SMC 
are now underway.  This will rectify one of the remaining short comings of 
the models -- the near order-of-magnitude discrepancy between the mass of the 
simulated and observed Stream.
These results represent the first self-consistent gas + N-body + star formation
simulations of the Magellanic System.

\end{article}
\end{document}